\documentclass[12pt]{article}
\usepackage[utf8]{inputenc}
\usepackage[margin=1in]{geometry}
\usepackage{amsmath}
\usepackage{bm}
\usepackage{amsfonts}
\usepackage{amssymb}
\usepackage{graphicx}
\usepackage{cite}
\usepackage{xcolor}
\usepackage{sectsty}
\sectionfont{\centering}
\usepackage{titlesec}
\usepackage{hyperref}
\hypersetup{
	colorlinks=true,
	linkcolor=blue,
	filecolor=magenta,
	urlcolor=cyan,
	citecolor=green,
	pdfpagemode=FullScreen
}
\usepackage{setspace}
\date{}
\usepackage{authblk}

\title{\large\textbf{Investigation of proton structure function $F_2 ^p$ at HERA in light of an analytical solution to Balitsky-Kovchegov equation} }

\author[1] {Ranjan Saikia 
\footnote{Corresponding author:  \href{mailto:ranjans@tezu.ernet.in}{ ranjans@tezu.ernet.in}}}

\author[2] {Pragyan Phukan \footnote{E-mail: \href{mailto:phukan.pragyan@gmail.com}{ phukan.pragyan@gmail.com}}}

\author[1] {Jayanta Kumar Sarma \footnote{E-mail: \href{mailto:jks@tezu.ernet.in}{ jks@tezu.ernet.in}}}

\affil[1]{\small \emph{HEP laboratory, Department of Physics, Tezpur University, Tezpur, Assam-784028, India}}

\affil[2]{\small \emph{Department of Physics, Moran College, Moranhat, Assam-785670, India}}

\begin{document}
\maketitle

\begin{abstract}
  In this paper, the proton structure function $F_2 ^p (x,Q^2)$ at small-$x$ is investigated using an analytical solution of the Balitsky-Kovchegov (BK) equation. In the context of the color dipole description of deep inelastic scattering (DIS), the structure function $F_2 ^p (x,Q^2)$ is computed by applying the analytical expression for the scattering amplitude $N(k,Y)$ derived from the BK solution. At transverse momentum $k$ and total rapidity $Y$, the scattering amplitude $N(k,Y)$ represents the propagation of the quark-antiquark dipole in the color dipole description of DIS. Using the BK solution we extracted the integrated gluon density $xg(x,Q^2)$ and then compared our theoretical estimation with the LHAPDF global data fits NNPDF3.1sx and CT18. Finally, we investigated the behavior of $F_2 ^p (x,Q^2)$ in the kinematic region of $10^{-5} \leq x \leq 10^{-2}$ and $2.5$ $GeV^{2}$ $\leq$ $Q^2$ $\leq$ $60$ $GeV^{2}$. Our predicted results for $F_2 ^p (x,Q^2)$ within the specified kinematic region are in good agreement with the recent high-precision data for $F_2 ^p (x,Q^2)$ from HERA (H1 Collaboration) and the LHAPDF global parametrization group NNPDF3.1sx. 
	
	\vspace{1cm}
	
    {\bf Keywords:} proton structure function, small-x, Balitsky-Kovchegov equation.
\end{abstract}
	
\thispagestyle{empty}

\clearpage
\section{Introduction}
\label{sec:intro}
Understanding the substructure of the nucleon within matter is one of the fundamental research topics in high-energy particle physics. It is crucial to understand the structure of the nucleon in order to comprehend the fundamental structure of matter. With the introduction of high-energy accelerator facilities, we have been able to understand the substructure of the nucleon within the framework of quantum chromodynamics (QCD). Understanding the substructure of the nucleon has relied heavily on the structure functions of the nucleon. Deep inelastic scattering (DIS) experiments on hadrons by leptons have yielded significant data on the distribution of partons within hadrons in terms of the quark and gluon distributions. The DIS cross section is associated with the structure functions of the nucleon in relation to parton distributions. The measurements of the proton's structure functions $F_2 ^p (x, Q^2)$ and $F_L ^p (x, Q^2)$ at HERA have begun a new era of parton density measurement within the nucleon~\cite{1,2,3,4}. These structure functions can be correlated with the momentum distributions of partons within the nucleon, and thus the parton distribution functions (PDFs). At high energies, or equivalently at small-$x$ (Bjorken $x$) values, the gluon density dominates among partons, and hence the dominant contribution in $F_2 ^p (x, Q^2)$ and $F_L ^p (x, Q^2)$ observations comes exclusively from gluons. Therefore, measurements of these structure functions at small-$x$ are essential for calculating gluon distribution functions and visualising the overall hadronic wave function in high-density QCD.

The DGLAP (Dokshitzer–Gribov–Lipatov–Altarelli–Parisi) evolution equation~\cite{a,b,c,d,e}, being the famous and well-established evolution equation, acts as a basic tool for the theoretical investigation of DIS structure functions. This equation has been successfully applied to address available HERA data in the moderate kinematical region of Bjorken's $x$ ($x$ $\geq$ $0.01$). At small-$x$, HERA data shows a steep rise in the behaviour of gluons in the region. This steep behaviour of gluons at small-$x$ is well described by the famous BFKL (Balitsky-Fadin-Kuraev-Lipatov) evolution equation~\cite{f,g}. However, the rapid growth of gluons at small-$x$ cannot continue indefinitely; otherwise, the physical cross-sections will violate unitarity and the Froissart-Martin bound~\cite{h}. As a result, both the DGLAP and BFKL evolution equations fail to account for implicit physics in high-density QCD. In order to maintain the unitarity of the theory, the infinite growth of gluons must be slowed down by certain processes. The phenomena of gluon recombination and saturation have provided the solution to the problems faced at small-$x$ by the linear evolution equations~\cite{i,j,k,l,m}. The nonlinear phenomena of gluon recombination and saturation lead to nonlinear terms in the DGLAP and BFKL equations. The Balitsky-Kovchegov (BK) equation~\cite{n,o,p,q}, nonlinearization of the BFKL equation, and mean-field approximation to the Jalilian-Marian-Iancu-McLerran-Weigert-Leonidov-Kovner (JIMWLK) equation~\cite{r,s,t,u} describe the behaviour of gluon density in the small-$x$ region. The BK equation is an integro-differential equation in coordinate space that can be transformed to momentum space to yield a partial differential equation useful for phenomenological studies at various high-energy accelerator facilities. This equation has been studied and solved numerically; however, due to its complexity, we have not seen an exact analytical solution to this equation to date. We recently proposed an approximate analytical solution to the BK equation that can be useful for phenomenology in the context of gluon saturation and hence small-$x$ physics~\cite{v}.

The collaborations at HERA contributed high-precision data for measuring the proton structure function $F_2 ^p$ at different kinematical regions of $x$ and $Q^2$~\cite{w,x}. Some recent works on the measurement of proton structure functions using various evolution equations and approaches are described in refs.~\cite{y,y1,z,aa,ab}, and the results correlate well with the experimental data. In ref.~\cite{ac}, for the first time, the reproduction of measurements of the DIS proton structure function at high energy from the color dipole description in momentum space was investigated. Using the knowledge of asymptotic solutions of the BK equation, the authors of ref.~\cite{ac} measured the charm structure function $F_2 ^c$ as a function of $x$ for various values of $Q^2$ and compared their results favourably with those of the HERA experiment. At small-$x$, it has been shown at HERA that the contributions in PDFs comes exclusively from gluons. Thus, the determination of gluon density is essential in order to understand the overall hadronic wave function at small-$x$. We do not see any theoretical measurement of gluon density at small-$x$ using BK evolution theory, which motivates the current work. We previously presented an analytical expression for the scattering amplitude $N(k, Y)$, which describes the propagation of the quark-antiquark dipole through the target color field in the color dipole model at various rapidities, $Y$~\cite{v}. In this work, we want to extract the gluon density $xg(x,Q^2)$ and analyze the proton structure function $F_2 ^p$ at small-$x$ by combining experimental data with QCD evolution theory using the analytical solution of the BK equation in momentum space. We compute the gluon density $xg(x,Q^2)$ and the proton structure function $F_2 ^p$ at various kinematical regions of HERA and analyze their behavior at small-$x$. 

The plan of the paper is as follows: In Section \ref{sec:color}, we relate the dipole-proton cross-section with forward scattering amplitude within the color dipole description of DIS in QCD. Then, using the analytical solution of the BK equation, we discuss how the structure function $F_2 ^p$ of the proton can be obtained from the color dipole description. In Section \ref{sec:results}, we present the $x$ evolution of the integrated gluon density $xg(x,Q^2)$ for two $Q^2$ values, viz. $35$ $GeV^2$ and $100$ $GeV^2$ using the BK evolution theory. Our results are compared with the LHAPDF global data fits NNPDF3.1sx~\cite{15} and CT18~\cite{16}. Both have LHC and HERA data included in their analyses. Finally, the obtained numerical results on the proton structure function $F_2 ^p$ using the analytical solution of the BK equation and comparison with the data from the H1 Collaboration and LHAPDF global parameterization group NNPDF3.1sx~\cite{15} are shown. Section \ref{sec:con} follows the discussion of the summary and conclusion of our work.

\section{Proton structure functions in the color dipole description}\label{sec:color}

\subsection{Color dipole description}
In QCD, the color dipole description of DIS~\cite{ad,ae,af} has been a useful tool for various perturbative QCD (pQCD) calculations at small-$x$ of high density QCD. It is especially convenient to work with the pQCD dipole description of DIS when it comes to the energy dependence of the scattering amplitude of any DIS event. The scattering process in the color dipole description of DIS can be factored into several steps where an incoming high-speed virtual photon ($\gamma^{\ast}$) after fluctuation from QCD vacuum changes to a quark-antiquark ($q\bar{q}$) dipole. The formed dipole then scatters off the target proton, eventually forming the final-state particles. In the color dipole description, following ref.~\cite{ag}, the total $\gamma^{\ast}p$ cross section for the $\gamma^{\ast}p$ scattering can be factored as
\begin{equation}
	\sigma^{\gamma^{\ast}p}_{T,L} =  \int d^2 r \int_{0} ^{1} dz |\Psi_{T,L}(r,z,Q^2)|^2 \sigma_{q\bar{q}}(r, Y) ,
\end{equation}
$\Psi_{T,L}$ represents the wave function for the virtual photon ($\gamma^{\ast}$) to fluctuate into the quark-antiquark ($q\bar{q}$) dipole with a transverse size of $r$. $T$, $L$ represent the transverse and longitudinal polarisation states of the virtual photon, respectively, and $z$ represents the longitudinal momentum fraction of the virtual photon carried by the quark (or antiquark). $Q^2$ represents the virtuality of the photon, while $Y$ represents the total rapidity. The dipole-proton cross section $\sigma_{q\bar{q}}(r, Y)$, which is the imaginary component of the forward scattering amplitude of the dipole on the target proton, contains all the information on the hadronic interactions.

The dipole-proton cross section, $\sigma_{q\bar{q}}(r, Y)$, can be obtained from the elastic dipole-proton scattering amplitude $\mathcal{A}(r, Y)$ using the optical theorem as~\cite{ah,ai}
\begin{equation}
	\sigma_{q\bar{q}}(r, Y) = 2 \hspace{1mm}\text{Im} \mathcal{A}(r, Y)=2\int d^2 b N(r,Y,b)=\sigma_0 N(r,Y).
\end{equation}
$N(r, Y, b)$ is the value for the imaginary part of the forward elastic dipole-proton scattering amplitude. Now, if one considers the target proton as a homogeneous disk of radius $R_p$, then the dipole-proton cross section can be related to the forward scattering amplitude $N(r, Y)$ by the following relation~\cite{ac} :
\begin{equation}
	\sigma_{q\bar{q}} (r,Y) = 2\pi R_p ^2 N (r,Y) .
\end{equation} 
The scattering amplitude $N(r,Y)$ will come from the solution of the BK equation.

After obtaining the $\gamma^{\ast}p$ cross-section, one can obtain directly the proton structure function $F_2 ^p$ from the $\gamma^{\ast}p$ cross-section through the relation
\begin{equation}
	F_2 ^p (x,Q^2) = \frac{Q^2}{4 \pi^2 \alpha_{em}} \left[\sigma^{\gamma^{\ast}p} _T (x, Q^2) + \sigma^{\gamma^{\ast}p} _L (x, Q^2)\right].
\end{equation}

\subsection{$F_2 ^p$ from the solution of the BK equation}
Now, we see how the structure function $F_2 ^ p$ can be obtained using the scattering amplitude $N(r,Y)$ obtained from the analytical solution of the BK equation. The BK equation describes the high-energy evolution of the dipole-target proton scattering amplitude $N(r,Y)$ in the dipole description, which in turn describes the propagation of the quark-antiquark dipole through the target color field. We express the $\gamma^{\ast}p$ cross-section in $N(k,Y)$ and hence the  structure function of the proton in momentum space, as the asymptotic behaviour of the solution of the BK equation can be expressed naturally in momentum space. For that, let us transform $N(r,Y)$ to $N(k,Y)$ by the following simple Fourier transform 
\begin{equation}
	N(k,Y) = \frac{1}{2\pi}\int \frac{d^2 r}{r^2} e^{i{k.r}} N(r,Y) = \int_{0} ^{\infty} \frac{dr}{r} J_0 (kr) N(r,Y).
\end{equation}

The proton structure function $F_2 ^p$ in momentum space related to $N(k,Y)$ can be expressed as follows using the discussion from above and some algebraic calculations~\cite{ac}: 
\begin{equation}\label{eqn:6}
	F_2 ^p (x, Q^2) = \frac{Q^2 R_p ^2 N_c}{4\pi^2} \int_{0}^{\infty}\frac{dk}{k}\int_{0}^{1} dz |\tilde{\Psi} (k^2 , z; Q^2)|^2 N(k,Y) ,
\end{equation}
where the photon wave function $\tilde{\Psi}$ is now expressed in the momentum space and can be found in ref.~\cite{ac} and $N_c$ are the number of colors. The scattering amplitude $N(k, Y)$ comes from the solution of the BK equation. The scattreing amplitude $N(k,Y)$ at total rapidity $Y$ and transverse momentum $k$ obeys the BK equation in momentum space as~\cite{p}
\begin{equation}\label{eqn:7}
	\partial_Y N = \bar{{\alpha}}\chi(-\partial_L)N - \bar{{\alpha}}N^2 ,
\end{equation}
where $\bar{{\alpha}} = \frac{\alpha_s N_c}{\pi}$ and $\chi (\gamma) = 2\psi(1)-\psi(\gamma)-\psi(1-\gamma)$ is the BFKL kernel, $\gamma = -\partial_L$, $L=\ln\frac{k^2}{k_0 ^2}$ with $k_0$ being some fixed low momentum scale. When expanding $\chi (\gamma)$ to second order, it is seen~\cite{11,12,13} that with some variable transformation the BK equation can be rewritten as the FKPP (Fisher-Kolmogorov-Petrovsky-Piscounov) equation as~\cite{aj,ak}
\begin{equation}\label{eqn:8}
	\partial_t u(t,x) = \partial_x ^2 u(t,x) + u(t,x) -u^2 (t,x) .
\end{equation}
Both the equations belong to the same universality class, and their solutions are of the same nature, i.e., travelling wave nature. We suggested an approximate analytical solution of the BK equation \eqref{eqn:7} in connection with the FKPP equation \eqref{eqn:8} using the homotopy perturbation method (HPM) in ref.~\cite{v}. Following ref.~\cite{v}, the solution of the BK equation \eqref{eqn:7} in connection with the FKPP equation \eqref{eqn:8} is given by
\begin{equation}\label{a}
	N(k,Y) = \frac{Ne^Y}{1-N+Ne^Y}.
\end{equation}
$N$ is the initial condition at $Y=0$ i.e. $N(k,0)=N$. This solution of the BK equation gives the scattering amplitude $N(k,Y)$ at any given rapidity $Y>0$, once the initial condition $N$ is known to us. For the initial condition $N$, we use the GBW (Golec-Biernat and Wusthoff) initial condition given by~\cite{14}
\begin{equation}\label{b}
	N (r,Y=0)=1-\text{exp}\left[-\left(\frac{r^2 Q_{s0} ^2}{4}\right)\right].
\end{equation}
$Q_{s0} ^2$ is the initial saturation momentum of gluons that can be fitted from the existing HERA data, and its value is $0.24$ $GeV^{2}$~\cite{al}. As we are dealing with the the BK equation in momentum space, the use of the GBW initial condition would be helpful as this can be transfomred into momentum space simply results as
\begin{equation}
	N(k,Y=0) = \int \frac{d^2 r}{2\pi r^2} e^{ik.r} N(r,Y=0) =\frac{1}{2}\Gamma\left(0, \frac{k^2}{Q_{s0} ^2}\right).
\end{equation}
At large values of $k^2 / Q_{s0} ^2$, the incomplete gamma function $\Gamma(0,k^2/Q_{s0} ^2)$ behaves as 
\begin{equation}
	\Gamma\left(0,\frac{k^2}{Q_{s0} ^2}\right)=\text{exp}\left(-\frac{k^2}{Q_{s0} ^2}\right).
\end{equation}
Therefore, we can write the GBW initial condition \eqref{b} in momentum space as 
\begin{equation}
	N(k,Y=0) =\frac{1}{2}\text{exp}\left(-\frac{k^2}{Q_{s0} ^2}\right).
\end{equation} 
Now, we use the above equation in equation \eqref{a} and replace the initial condition $N$ with the GBW initial condition in momentum space, which gives
\begin{equation}\label{c}
	N(k,Y)=\frac{e^{Y-k^2/Q_{s0} ^2}}{1-e^{-k^2/Q_{s0} ^2}+e^{Y-k^2/Q_{s0} ^2}}.
\end{equation}  
The scattering amplitude expression given above is an approximate analytical solution to the BK equation \eqref{eqn:7}. 

In light of the above discussions, we will now investigate the proton structure function $F_2 ^p$ by combining experimental data from the HERA experiment with QCD evolution theory. To do this, we will use the analytical solution of the BK equation, which is expressed above.

\section{Results and discussions} \label{sec:results}
\begin{figure}[h]
	\centering
	\includegraphics[width=0.6\linewidth]{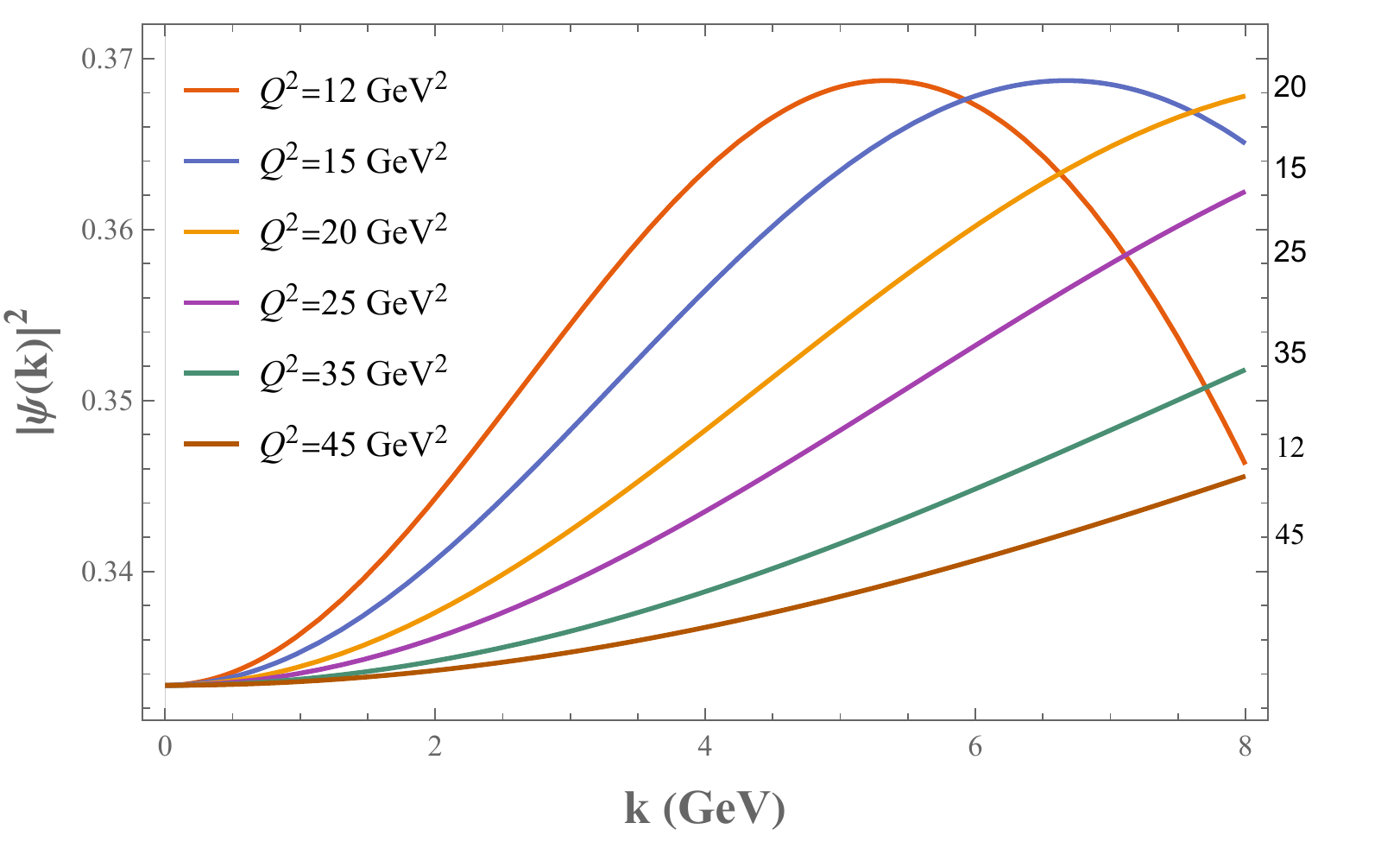}
	\caption{\label{fig:1} The probability distribution of the virtual photon to emit a quark-antiquark pair as a function of $k$ with $m_q \to 0$ and $z = \frac{1}{2}$ for various $Q^2$.}
\end{figure}

To investigate the proton structure function $F_2 ^p$, we use the expression eqn.\eqref{eqn:6} with the analytical solution of the BK equation eqn.\eqref{c}. We handle the expression eqn.\eqref{eqn:6} numerically for various inputs of $x$ and $Q^2$. In eqn.\eqref{eqn:6}, $\tilde{\Psi} (k^2 , z; Q^2)$ represents the probability of a virtual photon emitting a quark-antiquark pair with the momentum fraction $z$ (quark) and ($1-z$) (antiquark) of the virtual photon in the momentum space. Its expression is given by~\cite{ac}
\begin{equation}\label{eqn:10}
	\begin{aligned}
		|\tilde{\Psi}(k^2,z;Q^2)|^2 & = \sum_{q} \left(\frac{4\bar{Q}^2 _q}{k^2 + 4\bar{Q}^2 _q}\right) e^2 _q \Biggl\{[z^2 + (1-z)^2]\\
		& \times \Biggr[\frac{4(k^2+\bar{Q}^2 _q)}{k^2\sqrt{k^2 + 4\bar{Q}^2 _q}} \hspace{1mm}\text{arcsinh}\left(\frac{k}{2\bar{Q}_q}\right)+\frac{k^2 - 2\bar{Q}^2 _q}{2\bar{Q}^2 _q}\Biggr]\\
		& + \frac{4Q^2 z^2 (1-z)^2 + m^2 _q}{\bar{Q}^2 _q} \times \Biggr[\frac{k^2 + \bar{Q}^2 _q}{\bar{Q}^2 _q}-\frac{4\bar{Q}^4 _q + 2\bar{Q}^2 _q k^2 + k^4}{\bar{Q}^2 _q \sqrt{k^2 (k^2 + 4\bar{Q}^2 _q)}} \times \text{arcsinh}\left(\frac{k}{2\bar{Q}_q}\right)
		\Biggr] \Biggl\},
	\end{aligned}
\end{equation}
where $\bar{Q}_q ^2 = z(1-z)Q^2 + m_q ^2$ and $m_q$ is the mass of the quark of flavor $q$. Now, we see how the expression eqn.\eqref{eqn:10} will behave with transverse momentum $k$ for differenet values of $Q^2$ with $m_q \to 0$ and $z = 1/2$ (as $0 < z < 1$). The probability distribution graph of the virtual photon to emit a quark-antiquark pair with $z=1/2$ against transverse momentum $k$ is depicted in Figure \ref{fig:1}.
The probability to emit a quark-antiquark pair increases as $k$ increases until a certain point (peak point), after which it decreases as $k$ increases. The peak point for every $Q^2$ value is the same, but the transverse momentum $k$ corresponding to every peak is different for different $Q^2$.

\begin{figure}[h]
	\centering
	\begin{minipage}{.5\textwidth}
		\centering
		\includegraphics[width=1\linewidth]{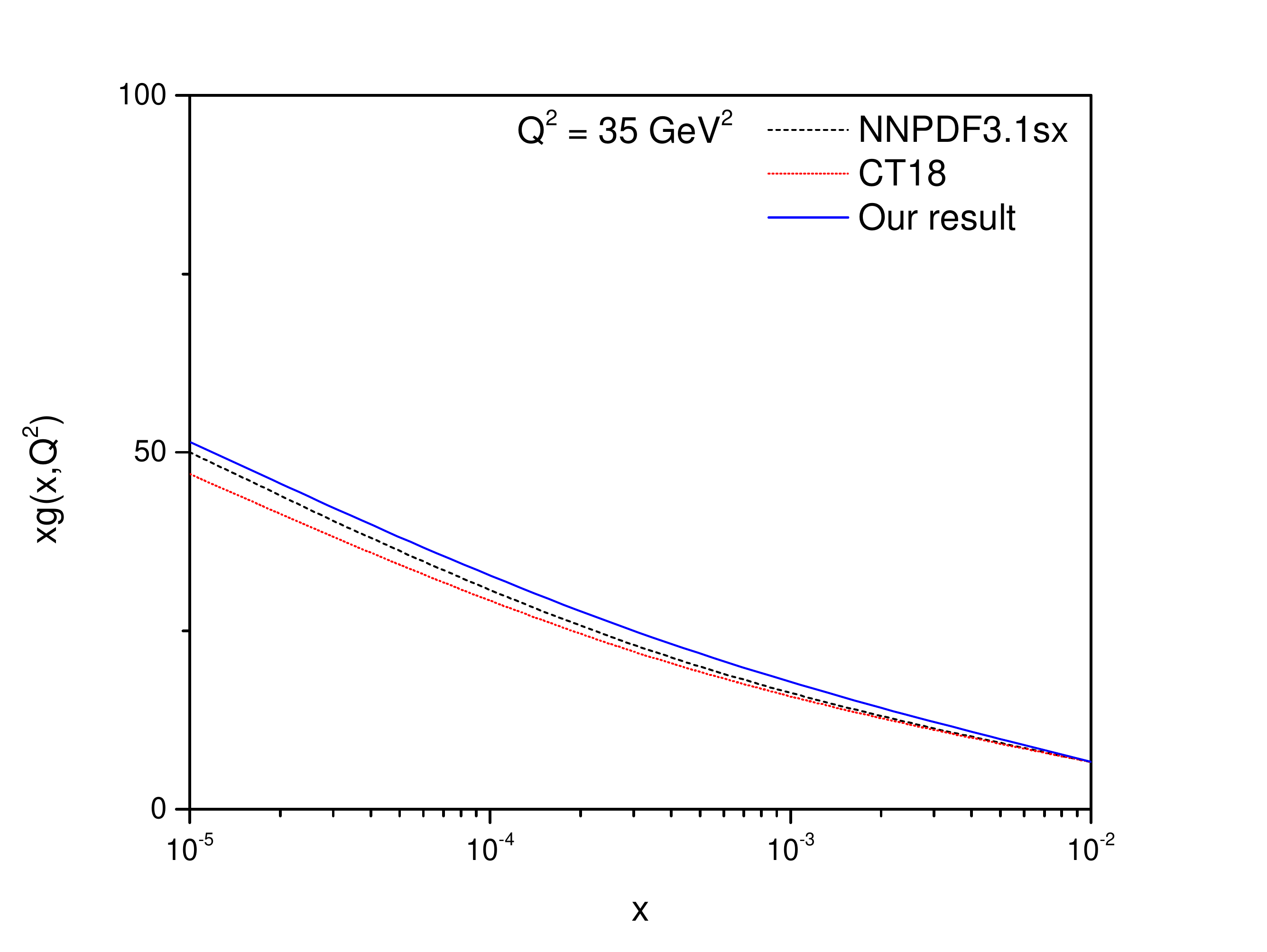}
		\label{fig:test1}
	\end{minipage}%
	\begin{minipage}{.5\textwidth}
		\centering
		\includegraphics[width=1\linewidth]{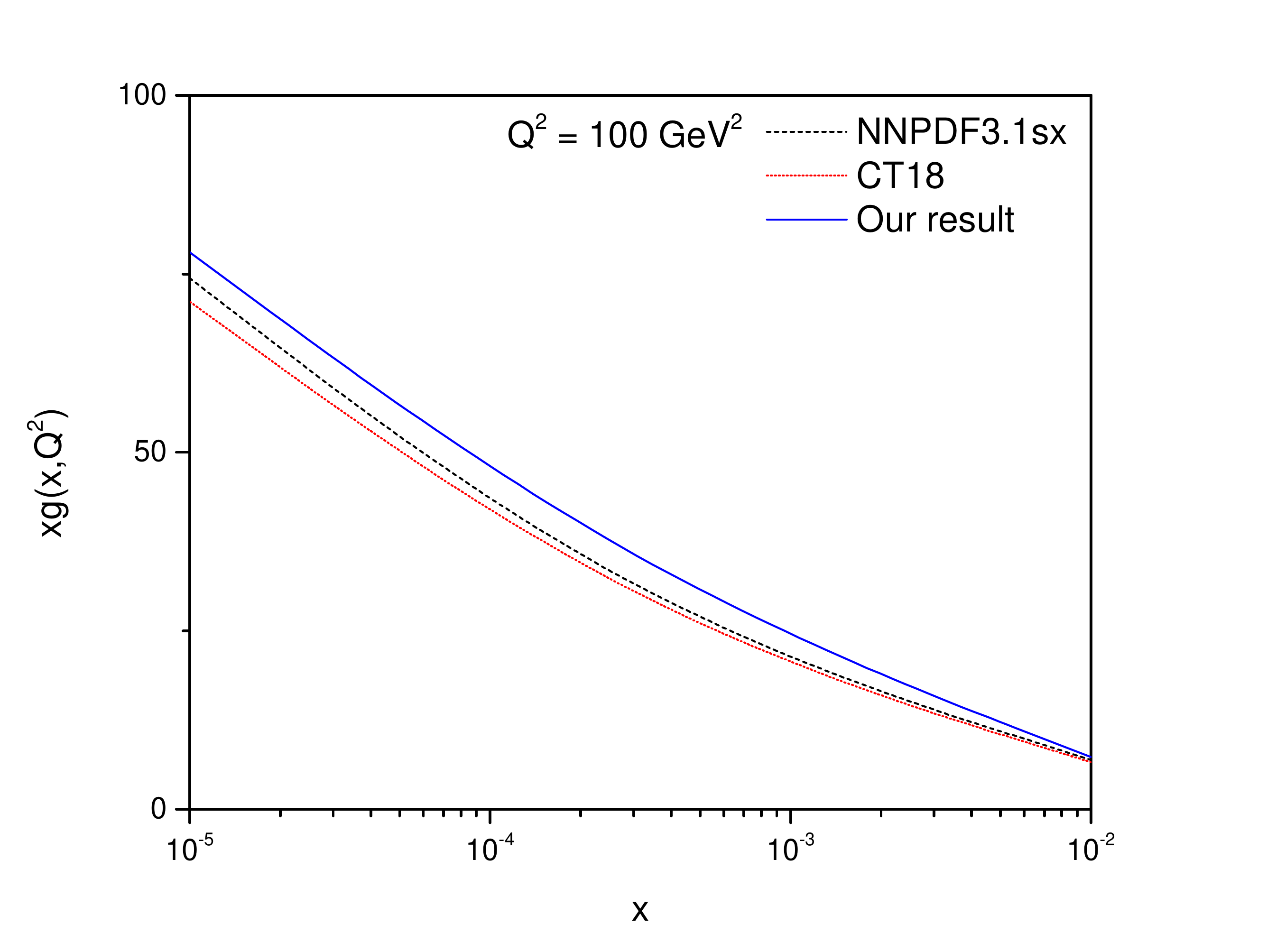}
		\label{fig:test2}
	\end{minipage}
	\caption{\label{fig.2}{$x$ evolution of gluon density $xg(x,Q^2)$ extracted from the BK solution compared with the global data fits NNPDF3.1sx and CT18. }}
\end{figure}

The solution \eqref{c} gives the propagation of the $q\bar{q}$ dipole in the color dipole description of QCD, which in turn gives the gluon content of the proton, i.e., the unintegrated gluon density. To calculate conventional gluon density $xg(x,Q^2)$, we integrate the BK solution \eqref{c} over transverse momentum with the relation
\begin{equation}\label{d}
	xg(x,Q^2) = \int_0 ^{Q^2} \frac{dk}{k}N(k,Y).
\end{equation}
The extracted results of gluon density from BK solution is compared with the global data fits NNPDF3.1sx and CT18. The results are shown in Figure \ref{fig.2}.

\begin{figure}[h]
	\centering
	\includegraphics[width=1.2\linewidth]{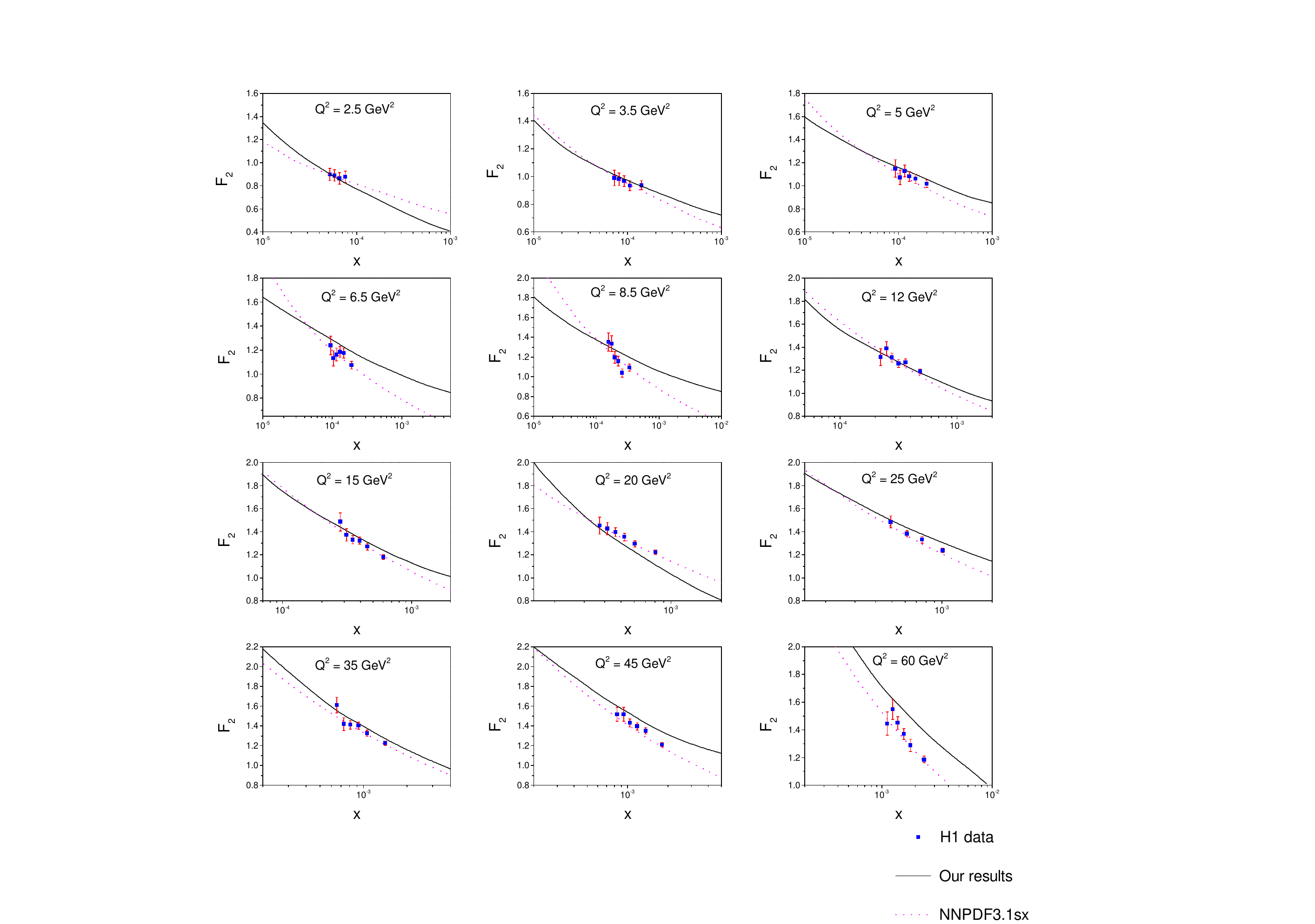}
	\caption{\label{fig:3}The results of the proton structure function $F_2 ^p$ as a function of $x$ at various $Q^2$ obtained in this work are compared with the data from the H1 Collaboration~\cite{x} and global data fit NNPDF3.1sx~\cite{15}.}
\end{figure}

For the proton structure function $F_2 ^p$, we plug the expression given in eqn.\eqref{eqn:10} (with $m_q \to 0$ and $z=1/2$) together with the BK solution given in eqn.\eqref{c} into eqn.\eqref{eqn:6}. We set $N_c = 3$ and $R_p \approx 0.831$ $fm$ $\approx 4.22$ $GeV^{-1}$ in the expression eqn.\eqref{eqn:6} and solve the expression numerically for various $Q^2$. The value of the $R_p$ is taken from the recent study on the proton radius~\cite{am}. We analyse the expression eqn.\eqref{eqn:6} numerically and compare our results with the HERA measurements of the proton structure function from the H1 Collaboration~\cite{x} with constraints to the kinematic region: $10^{-5} \leq x \leq 10^{2}$ and $2.5 \leq Q^2 \leq 60$ $GeV^{2}$. We also compare our results with the LHAPDF global parameterization group NNPDF3.1sx~\cite{15}. The reason for choosing the particular kinematic region $x\leq 10^{-2}$ is to describe the small-$x$ behaviour of high-energy amplitude as the BK equation is only applicable at small-$x$. For the too high $Q^2$ range, we would need corrections from the DGLAP equation, which we cannot skip at too high $Q^2$. The results are shown in Figure \ref{fig:3}.

\section{Summary and conclusion}\label{sec:con}
In this work, we have investigated the proton structure function $F_2 ^p$ in light of an analytical solution of the BK equation. Proton structure functions have been investigated by different collaborations at experimental facilities such as HERA and the LHC. On the phenomenological side, the proton structure functions have been studied by studying different QCD evolution equations and tested well at existing experimental facilities. To investigate the proton structure functions at small-$x$, the BK equation is most suitable for testing the experimental data on observables. In this work, we have stated our discussion on how we can obtain the proton structure function $F_2 ^p$ from the color dipole description of the DIS in QCD. In the color dipole approach, the proton structure function $F_2 ^p$ is obtained directly from the virtual photon-proton cross-section, which can be expressed in the dipole-proton scattering amplitude $N(k,Y)$. The dipole-proton scattering amplitude $N(k,Y)$ is obtained from the solution of the BK equation. Using the BK solution, we extracted the $x$ evolution of the integrated gluon density $xg(x,Q^2)$ and compared our results with those of LHAPDF global data fits NNPDF3.1sx and CT18. Both NNPDF3.1sx and CT18 have HERA and LHC data included in their analyses. Finally, we computed the proton structure function $F_2 ^p$ and compared it with the measurement of $F_2 ^p$ at HERA from the H1 Collaboration and the LHAPDF global parameterization group NNPDF3.1sx. Our predicted results are in good agreement with the experimental results within the kinematic region we have constrained. Beyond that, we have to consider corrections to the DGLAP equation. We have also shown in this work how the virtual photon wave function $\Psi$ would behave against transverse momentum $k$ with $m_q \to 0$, $z=1/2$ at various $Q^2$. We found that the probability of the virtual photon emitting a quark-antiquark pair increases as $k$ increases to a certain peak value, after which it starts to fall as $k$ increases. We have seen that the maximum probability is the same for different $Q^2$ values, irrespective of the value of $k$ at the maximum peak for different $Q^2$ values.

In this work, we have seen the ability of our BK solution to describe physics at small-$x$. We have successfully applied our BK solution to calculate the gluon density $xg(x,Q^2)$ and the proton's structure function $F_2 ^p$ at small-$x$ within the kinematic region we have constraint. Also, our results are testable at future experimental facilities such as the LHeC (Large Hadron electron Collider)~\cite{an,ao}, EIC (Electron Ion Collider)~\cite{ap}, and the FCC-eh (Future Circular Collider electron-hadron)~\cite{aq}. In these future experimental facilities, the measurement of the proton structure function will be performed at much lower values of $x$ with increased precision. Nevertheless, we could investigate the proton structure function $F_2 ^p$ using an analytical solution of the BK equation within the constrained region. We conclude that the analytical solution of the BK equation can serve as a convenient tool for further studies at small-$x$ and high-density QCD. We hope that the BK equation, together with future experimental facilities, will help us understand and explore phenomena inside hadrons at small-$x$ in the near future.

\begin{center}
\rule{8cm}{0.3mm}
\end{center}

\end{document}